\documentclass[10pt, conference]{IEEEtran}
\IEEEoverridecommandlockouts
\usepackage{cite}
\usepackage{amsmath,amssymb,amsfonts}
\usepackage[ruled,vlined]{algorithm2e}

\usepackage{tikz}
\usepackage{algpseudocode}
\usepackage{graphicx}
\usepackage{textcomp}
\usepackage{xcolor}
\usepackage{color}
\usepackage{makecell}
\usepackage{amsmath, amssymb, amsthm}
\usepackage{subcaption}
\usepackage{caption} 
\usepackage{setspace} 
\usepackage{pdfpages}
\newtheorem{theorem}{Theorem}
\newtheorem{assumption}{Assumption}
\newtheorem{remark}{Remark}
\newtheorem{definition}{Definition}

\def\BibTeX{{\rm B\kern-.05em{\sc i\kern-.025em b}\kern-.08em
    T\kern-.1667em\lower.7ex\hbox{E}\kern-.125emX}}
\begin{document}

% \title{Over-the-Air Federated Learning with Age-Aware Partial Gradient Updates\\
% % {\footnotesize \textsuperscript{*}Note: Sub-titles are not captured in Xplore and
% % should not be used}
% % \thanks{Identify applicable funding agency here. If none, delete this.}
% }

\title{Age-Aware Partial Gradient Update Strategy for Federated Learning Over the Air\\
% \thanks{*Corresponding Author: Howard H. Yang.}
}

% \author{\IEEEauthorblockN{ Ruihao Du}
% \IEEEauthorblockA{\textit{ZJU-UIUC Institute} \\
% \textit{Zhejiang University}\\
% Haining, China \\
% ruihao.24@intl.zju.edu.cn}
% \and
% \IEEEauthorblockN{ Zeshen Li}
% \IEEEauthorblockA{\textit{ZJU-UIUC Institute} \\
% \textit{Zhejiang University}\\
% Haining, China \\
% zeshen.22@intl.zju.edu.cn}
% \and
% \IEEEauthorblockN{ Howard H. Yang*}
% \IEEEauthorblockA{\textit{ZJU-UIUC Institute} \\
% \textit{Zhejiang University}\\
% Haining, China \\
% haoyang@intl.zju.edu.cn}

% }
\author{%
  \IEEEauthorblockN{
  Ruihao~Du,
  Jiaqi~Zhu,
  Zeshen~Li,
  and Howard~H.~Yang
  }
  \IEEEauthorblockA{
                    ZJU-UIUC Institute, Zhejiang University, Haining, China}
}

\maketitle

\begin{abstract}
Frequent parameter exchanges between clients and the edge server incur substantial communication overhead, posing a critical bottleneck in federated learning (FL).
By exploiting the superposition property of wireless waveforms, over-the-air (OTA) computation enables simultaneous analog aggregation of local updates, thereby reducing communication latency and improving spectrum efficiency.
However, its scalability is constrained by the limited number of available orthogonal waveform resources, which are typically far fewer than the model dimension.
To address this, we propose AgeTop-$k$, an age-aware gradient sparsification strategy that performs compression through a two-stage selection process.
Specifically, the edge server first selects candidate gradient entries based on their magnitudes, and then further prioritizes them according to the Age of Information (AoI), which quantifies the staleness of updates.
AoI tracking is achieved efficiently by maintaining an age vector at the edge server.
We derive theoretical convergence guarantees for non-convex loss functions and demonstrate the efficacy of AgeTop-$k$ through extensive simulations.
\end{abstract}


\section{Introduction}

Federated learning (FL) enables multiple clients to collaboratively train a global model while preserving data privacy. 
However, frequent parameter exchanges for model updates between clients and the edge server introduce substantial communication overhead, which scales with the number of clients and model complexity, constituting the communication bottleneck in FL \cite{park2019wireless, yang20federated}.
To mitigate this issue, recent studies \cite{amiri20machine, sery20analog, yang20federated, zhu19broadband} have incorporated over-the-air (OTA) computation \cite{nazer07computation} into FL. 
By exploiting the superposition property of wireless waveforms, OTA enables simultaneous transmission and aggregation of model updates directly at the edge server, significantly improving spectral and energy efficiency while reducing overall communication latency \cite{yang21revisiting, chen23edge, yang24unleashing}.

Despite these advantages, a critical limitation for OTA-FL lies in the reliance on orthogonal waveforms.
% The received analog signals at the edge server are inevitably distorted by channel fading and thermal noise \cite{cao20optimized}. 
The limited number of available orthogonal waveforms is often far fewer than the model dimension \cite{sery20analog}, making gradient compression indispensable for scalable OTA-FL systems.
Several compression schemes have been proposed for OTA-FL \cite{amiri20machine, ahn22model, tao24private, zhang21federated, barnes20rtop, zheng25toward}.
For example, \cite{amiri20machine} combines the widely adopted Top-$k$ sparsification with random sketching to compress local gradients for analog transmission.
However, the conventional Top-$k$ method updates only the largest-magnitude gradients and the superposition of sparse local updates is often no longer sparse, leading to significant reconstruction errors, especially in large-scale client scenarios \cite{ahn22model}.
Furthermore, the error accumulation mechanism in \cite{amiri20machine} may defer the update of low-magnitude but important gradients, resulting in prolonged staleness for some parameters.
Alternatively, \cite{tao24private, zhang21federated} adopt Random-$k$ sparsification with a uniform pattern across all clients, ensuring strict alignment but lacking the ability to prioritize critical gradients.

Hybrid methods have also been proposed to balance gradient magnitude and random exploration.
For example, rTop-$k$ \cite{barnes20rtop} first identifies the top $r$ gradient entries with the largest magnitudes and subsequently randomly selects $k$ entries from this subset to reduce communication overhead.
TopRand \cite{zheng25toward} exploits temporal correlation across training rounds by deterministically selecting the largest gradients while randomly sampling the remaining entries to compensate for potential accumulation errors.
Nonetheless, the inherent randomness in these methods may cause some critical gradient coordinates to be unupdated for extended periods, resulting in stale updates and degraded convergence efficiency.

To address this issue, we maintain a global gradient vector at the edge server, updated based on the gradients received from clients in each communication round.
We then introduce the Age of Information (AoI) metric, defined as the time elapsed since the most recently received information was generated at its source, thereby quantifying its staleness \cite{yates21age, howard20age}.
The AoI value of each entry in the global gradient vector is recorded to quantify the time since its last update.
By jointly considering gradient magnitude and AoI in the selection process, the proposed method enhances the directionality of updates and ensures that stale yet important gradients are refreshed timely, thereby achieving fairer and more efficient model training.

\begin{figure*}[t!]
    \centering
    \includegraphics[width=1\textwidth]{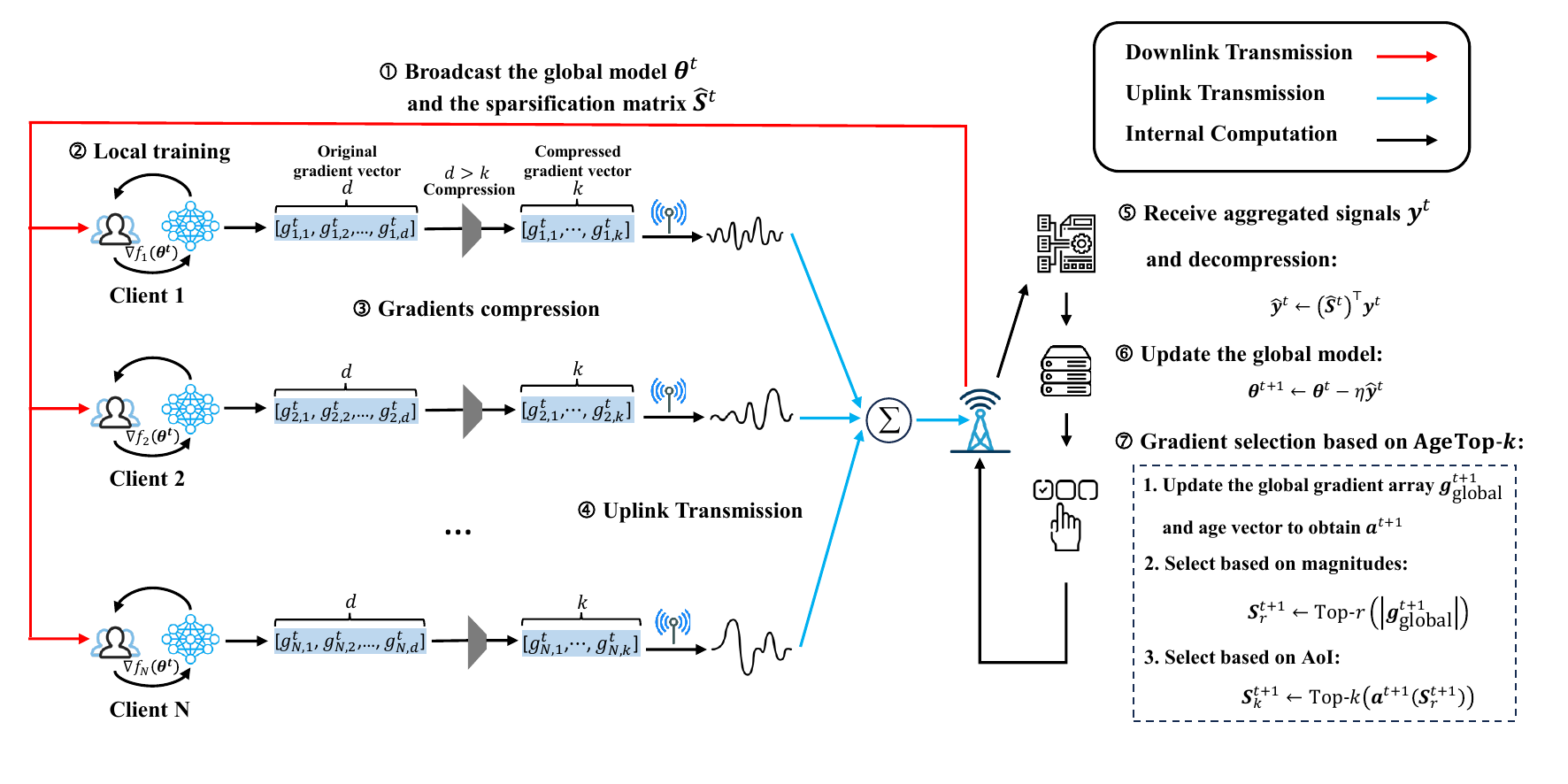} 
    \caption{An illustration of the considered system.}
    \label{fig:subfig-a}
\end{figure*}

The main contributions are summarized below.
\begin{itemize}
\item We propose AgeTop-$k$, an age-aware partial gradient update strategy for OTA-FL. The edge server compresses the gradients via a two-stage selection process, in which it identifies a candidate set of significant entries and then prioritizes them based on AoI, ensuring fairness and efficiency in model updates.
\item We theoretically derive the convergence rate of AgeTop-$k$, accounting for compression ratio, wireless channel distortion, and data heterogeneity, which provides insights into key factors affecting convergence rate and training performance.
\item We conduct extensive simulations using the Convolutional Neural Network (CNN) and ResNet-18 models on EMNIST and CIFAR-10 datasets and validate the effectiveness of AgeTop-$k$.
\end{itemize}

\section{System Model and Algorithm Design}
We consider a federated learning system consisting of $ N $ clients and an edge server. 
Each client $ n $ ($ n \in \{1, 2, \dots, N\} $) holds a local dataset $ \mathcal{D}_{n} = \{(\boldsymbol{x}_{i}, y_{i})\}_{i=1}^{m_n} $ with size $ |\mathcal{D}_{n}| = m_n $, where $\boldsymbol{x}_i \in \mathbb{R}^d$ and $y_i \in \mathbb{R}$ denote the input feature and its associated label, respectively. 
The global learning objective is formulated as
\begin{equation} \label{equ:GlbLosFunc}
\min_{\boldsymbol{\theta}} \; f(\boldsymbol{\theta}) = \frac{1}{N} \sum_{n=1}^{N} f_n(\boldsymbol{\theta}),
\end{equation}
where $ \boldsymbol{\theta} \in \mathbb{R}^d $ is the global model, and $ f_n(\boldsymbol{\theta}) $ is the local empirical loss defined as
\begin{equation}
f_n(\boldsymbol{\theta}) = \frac{1}{m_n} \sum_{i=1}^{m_n} \ell(\boldsymbol{\theta}; \boldsymbol{x}_{i}, y_{i}),
\end{equation}
with $ \ell(\cdot) $ denoting the loss evaluated on a single data sample.
The training process in \eqref{equ:GlbLosFunc} is carried out via OTA-FL.
At the beginning of each global iteration $t$, the edge server broadcasts the current global model $\boldsymbol{\theta}^{t}$ to all clients. 
Upon reception, each client $n$ computes its local gradient:
\begin{equation}
\boldsymbol{g}_n^{t} = \nabla f_n(\boldsymbol{\theta}^{t}).
\end{equation}
After local training,  each client obtains the gradient \( \boldsymbol{g}_n^{t} \), which is transmitted to the edge server using an analog transmission scheme.
Specifically, each client modulates the gradient entries onto the amplitudes of a set of mutually orthogonal waveforms and 
simultaneously transmits the resulting analog signals over the shared spectrum.

Owing to the superposition property of wireless waveforms, the client gradients are automatically aggregated at the radio front end of the edge server.
The edge server can pass the received radio signal through a bank of matched filters, with each tuned onto one of the waveforms, and extract the globally aggregated (but noisy) gradient, which is then used to update the model parameters.
The edge server then broadcasts the updated model to all clients for another round of local training. 
This iterative process is repeated until the global model converges. 

However, in pragmatic wireless systems, the number of available orthogonal waveforms is typically far fewer than the model dimension, rendering the transmission of full gradients infeasible.
Consequently, only a subset of gradient entries can be transmitted in each communication round, and their selection significantly affects both the convergence rate and overall training performance.
To mitigate this limitation, we propose a low-complexity yet effective age-aware update strategy that adaptively selects transmitted entries based on both the gradient magnitudes and their corresponding AoI, as detailed below.

\subsection{Age-Aware Partial Gradient Update}

In our proposed age-aware gradient update strategy, the edge server selects gradient indices and broadcasts the corresponding sparsification matrix for each communication round, while maintaining a global gradient vector $\boldsymbol{g}_{\text{global}}^{t} \in \mathbb{R}^d$, initialized to zero to track the most recent updates. 
The update mechanism, detailed later, refreshes this vector each round. 
The strategy is performed in two consecutive stages.

In the \textit{first stage}, the edge server identifies the most significant gradient entries based on their magnitudes. 
Specifically, the $r$ entries with the largest magnitudes are selected from $\boldsymbol{g}_{\text{global}}^{t}$:
\begin{equation}
\boldsymbol{S}_r^{t} = \text{Top-}r\!\left( \lvert \boldsymbol{g}_{\text{global}}^{t} \rvert \right),
\end{equation}
where $\boldsymbol{S}_r^{t}$ denotes the index set of the $r$ largest-magnitude entries.

In the \textit{second stage}, we introduce the AoI metric to quantify how long each gradient entry has remained unselected, with larger AoI values indicating older, potentially stale gradients. 
The edge server maintains an age vector $\boldsymbol{a}^{t} \in \mathbb{R}^{d}$, initialized to zero, where $\boldsymbol{a}^{t}[j]$ records the AoI of the $j$-th entry. 
Based on the current AoI values, the $k$ entries with the largest values are selected from the candidate set $\boldsymbol{S}_r^{t}$:
\begin{equation}
\boldsymbol{S}_k^{t} = \text{Top-}k\!\left( \boldsymbol{a}^{t}(\boldsymbol{S}_r^{t}) \right),
\end{equation}
where $\boldsymbol{a}^{t}(\boldsymbol{S}_r^{t})$ denotes the AoI values of the corresponding entries.
The selected indices $\boldsymbol{S}_k^{t}$ are represented by a diagonal binary sparsification matrix $\boldsymbol{S}^{t} \in \{0,1\}^{d \times d}$, whose diagonal elements $s_{i,i}$ indicate selection:
\begin{equation}
s_{i,i} =
\begin{cases}
1, & i \in \boldsymbol{S}_k^{t}, \\
0, & \text{otherwise}.
\end{cases}
\end{equation}
Removing all-zero rows yields the compact sparsification matrix $\hat{\boldsymbol{S}}^{t} \in \{0,1\}^{k \times d}$.

Through this two-stage selection process, the edge server effectively targets gradient entries that are both of large magnitude and have remained unupdated for extended periods, achieving a compression ratio of $k/d$ and enabling more efficient model training.

\begin{algorithm}[t!]
\caption{AgeTop-$k$ Algorithm}
\label{alg:agetopk_pseudo}
\SetAlgoLined
\SetKwInOut{Input}{Input}
\SetKwInOut{Initialize}{Initialize}

\KwIn{Initial global model $\boldsymbol{\theta}^0 \in \mathbb{R}^d$, 
age vector $\boldsymbol{a}^0 = \boldsymbol{0} \in \mathbb{R}^d$, 
compression parameters $r$ and $k$,
global gradient vector $ \boldsymbol{g}_{\text{global}}^{0}  = \boldsymbol{0} \in \mathbb{R}^d$ 
% the sparsification matrix $\hat{\boldsymbol{S}}^{0} \in \mathbb{R}^{k \times d}$
.}

\For{$t = 0,1,2,\dots,T-1$}{
    \text{Edge server broadcasts the global model $\boldsymbol{\theta}^{t}$ and }\\
    \text{sparsification matrix $\hat{\boldsymbol{S}}^{t}$ to all clients} \\    
    \For{\textit{client $n = 1,\dots,N$ in parallel}}{
        \text{Train model locally and update the gradient $\boldsymbol{g}_n^{t}$} \\
        \text{\makecell[l]{Apply compression and transmit the compressed\\
        gradient $\tilde{\boldsymbol{g}}_n^{t} \gets \hat{\boldsymbol{S}}^{t} \boldsymbol{g}_n^{t}$}}
    }
    
    \textit{\# Receive the aggregated signals} \\
    $\boldsymbol{y}^{t} \gets \frac{1}{N} \sum_{n=1}^{N} h_n^{t} \tilde{\boldsymbol{g}}_n^{t} + \boldsymbol{\xi}^{t}$ \\
    \textit{\# Reconstruct the gradient} \\
    $\hat{\boldsymbol{y}}^{t} \gets (\hat{\boldsymbol{S}}^t)^\top \boldsymbol{y}^{t}$
    
    \textit{\# Update the global model}\\
    $\boldsymbol{\theta}^{t+1} \gets \boldsymbol{\theta}^{t} - \eta \hat{\boldsymbol{y}}^{t}$
    
    \textit{\# Update global gradient vector and age vector}\\
    $\boldsymbol{g}_{\text{global}}^{t+1}[j] \gets
      \begin{cases}
      \hat{\boldsymbol{y}}^{t}[j], & j \in \boldsymbol{S}_k^{t} \\
      \boldsymbol{g}_{\text{global}}^{t}[j], & \text{otherwise}
      \end{cases}$\\
      
    $\boldsymbol{a}^{t+1}[j] \gets
      \begin{cases}
      0, & j \in \boldsymbol{S}_k^{t} \\
      \boldsymbol{a}^{t}[j] + 1, & \text{otherwise}
      \end{cases}$\\    
    
    \textit{\# Update the sparsification for next round}\\
    $\boldsymbol{S}_r^{t+1} \gets \text{Top-}r\big(\lvert \boldsymbol{g}_{\text{global}}^{t+1} \rvert\big)$ \quad \textit{\# Magnitude}\\
    $\boldsymbol{S}_k^{t+1} \gets \text{Top-}k\big(\boldsymbol{a}^{t+1}(\boldsymbol{S}_r^{t+1})\big)$ \quad \textit{\# AoI}\\
    \text{Construct $\hat{\boldsymbol{S}}^{t+1}$ from $\boldsymbol{S}_k^{t+1}$}\\
}
\end{algorithm}

\addtolength{\topmargin}{0.01in}
\subsection{Model Training Over the Air}

At the beginning of each communication round, the edge server broadcasts the current global model $\boldsymbol{\theta}^{t}$ and the sparsification matrix $\hat{\boldsymbol{S}}^{t}$ to all clients. 
Upon reception, each client performs local training to compute $\boldsymbol{g}_n^{t}$ and applies the sparsification matrix to obtain
\begin{equation}
\tilde{\boldsymbol{g}}_n^{t} = \hat{\boldsymbol{S}}^{t} \boldsymbol{g}_n^{t}.
\end{equation}

Each client then maps its compressed gradient $\tilde{\boldsymbol{g}}_n^{t}$ onto the amplitudes of a shared set of orthogonal waveforms in an entry-wise manner and simultaneously transmits the modulated analog signals to the edge server. 
Each client performs power control to compensate for large-scale path loss \cite{li06rss}, while small-scale fading remains unknown, and all clients are subject to a per-round maximum transmit power constraint.

Exploiting the superposition property of the wireless waveforms, the edge server receives a noisy, aggregated version of the compressed gradients:
\begin{equation}
\boldsymbol{y}^{t} = \frac{1}{N} \sum_{n=1}^{N} h_n^{t} \tilde{\boldsymbol{g}}_n^{t} + \boldsymbol{\xi}^{t},
\end{equation}
where $h_n^{t}$ denotes the channel fading experienced by client $n$ at round $t$, and $\boldsymbol{\xi}^{t} \in \mathbb{R}^{k}$ is a vector comprised of the thermal noise.
We model the channel fading as a random variable with mean $\mu_h$ and variance $\sigma_h^2$, independent and identically distributed (i.i.d.) across clients and communication rounds, while the thermal noise is modeled as additive white Gaussian noise (AWGN) with variance $\sigma_z^2$.

Upon reception, the edge server reconstructs the $d$-dimensional sparse global gradient $\hat{\boldsymbol{y}}^{t}$:
\begin{equation}
\hat{\boldsymbol{y}}^{t} = (\hat{\boldsymbol{S}}^{t})^\top \boldsymbol{y}^{t}
= \frac{1}{N} \sum_{n=1}^{N} h_n^{t} \boldsymbol{S}^{t} \boldsymbol{g}_n^{t} + (\hat{\boldsymbol{S}}^{t})^\top \boldsymbol{\xi}^{t}.
\end{equation}
The global model $\boldsymbol{\theta}^{t}$ is updated using $\hat{\boldsymbol{y}}^{t}$:
\begin{equation}
\boldsymbol{\theta}^{t+1} = \boldsymbol{\theta}^{t} - \eta \hat{\boldsymbol{y}}^{t}.
\end{equation}
The edge server then updates both the global gradient vector $\boldsymbol{g}_{\text{global}}^{t}$ and the age vector $\boldsymbol{a}^{t}$ for the next iteration as follows:
\begin{align}
\boldsymbol{g}_{\text{global}}^{t+1}[j] &=
\begin{cases}
\hat{\boldsymbol{y}}^{t}[j], & j \in \boldsymbol{S}_k^{t}, \\
\boldsymbol{g}_{\text{global}}^{t}[j], & \text{otherwise},
\end{cases}
\\
\boldsymbol{a}^{t+1}[j] &=
\begin{cases}
0, & j \in \boldsymbol{S}_k^{t}, \\
\boldsymbol{a}^{t}[j] + 1, & \text{otherwise}.
\end{cases}
\end{align}
% The updated pair $(\boldsymbol{g}_{\text{global}}^{t+1}, \boldsymbol{a}^{t+1})$ serves as the input for the \textit{Age-Aware Partial Gradient Update} in the subsequent communication round. 
An overview of the considered OTA-FL system is illustrated in Fig.~1, and the detailed procedural steps are summarized in Algorithm~1.

\section{Convergence Analysis}

In this section, we analyze the convergence of the proposed AgeTop-$k$ algorithm under the OTA-FL system.
Our analysis explicitly accounts for the impact of compression ratio, client data heterogeneity, and channel noise introduced by OTA aggregation.
We begin by introducing a standard definition and several commonly used assumptions.

\begin{definition}
An operator $C: \mathbb{R}^{d} \to \mathbb{R}^{d}$ is called a \emph{$\gamma$-approximate compressor} if there exists $\gamma \in (0, 1]$ such that
\begin{equation}
\mathbb{E}\big[\|C(\boldsymbol{x}) - \boldsymbol{x}\|_2^2\big] 
\le (1 - \gamma)\|\boldsymbol{x}\|_2^2, 
\quad \forall \boldsymbol{x} \in \mathbb{R}^{d}.
\end{equation}
\end{definition}

\begin{assumption}
The ratio between the largest and the $r$-th largest gradient magnitudes is bounded by a constant $\beta$ for all clients and iterations.
\end{assumption}

Definition~1 provides a standard formulation of compression operators \cite{stich18sparsified}. 
Inspired by the analysis in \cite{mortaheb24r}, under Assumption~1, the proposed AgeTop-$k$ strategy can be viewed as a $\gamma$-approximate compressor with
\begin{equation}
\gamma = \frac{k}{k + (r - k)\beta + (d - r)}.
\end{equation}
In this expression, $k$ denotes the number of transmitted gradient entries, while the denominator accounts for all potential entries weighted by their relative significance. 
Hence, $\gamma$ captures the effective information retention of AgeTop-$k$ and bounds the worst-case compression error.

\begin{assumption}\label{assm:L-smooth}
    \textit{
    The objective function $f: \mathbb{R}^{d} \rightarrow \mathbb{R}$ is $L$-smooth, i.e., for any  $\boldsymbol{\theta}, \boldsymbol{v} \in \mathbb{R}^{d}$, it is satisfied:
    \begin{align}
        &f(\boldsymbol{\theta}) \leq f(\boldsymbol{v}) + \langle \nabla f(\boldsymbol{v}), \boldsymbol{\theta}- \boldsymbol{v} \rangle + \frac{L}{2} \Vert \boldsymbol{\theta} - \boldsymbol{v}\Vert ^2
    \end{align}
    where $L$ is a positive constant.
    }
\end{assumption}

\begin{assumption}\label{assm:heterogeneity bound}
    \textit{
    There exists a constant $\sigma_g > 0$ such that the difference among local objective functions is bounded: 
    \begin{equation}
        \frac{1}{N}\sum^N_{n=1} \|\nabla f_n(\boldsymbol{\theta})-\nabla f(\boldsymbol{\theta})\|^2 \leq \sigma_g^2.
    \end{equation}
    }    
\end{assumption}

\begin{assumption}\label{assm:gradient bound}
    \textit{
    The average of the expected squared norm of the gradients of functions $f_n(\boldsymbol{\theta})$ is bounded, i.e., there exists a positive constant $G$ that
    \begin{equation}
        \frac{1}{N} \!\sum^N_{n=1} \mathbb{E}\big[\| \nabla f_n(\boldsymbol{\theta})\|^2_2 \big] \leq G^2.
    \end{equation}
    }     
\end{assumption}

Assumptions~2–4 are standard in the convergence analysis of federated optimization \cite{li19convergence, stich18sparsified}. 
Assumption~3 captures data heterogeneity across clients, where a larger $\sigma_g$ indicates more significant inter-client divergence. 
Assumption~4 holds naturally in OTA-FL since each client operates under the maximum transmit power constraint.

We are now ready to present the convergence rate.

\begin{theorem}
\label{thm:convergence}
Under the considered OTA-FL system, the global model converges as
{\small
\begin{equation}
\frac{1}{T}\sum_{t=0}^{T-1}\mathbb{E}\!\left[\|\nabla f(\boldsymbol{\theta}^{t})\|_2^2\right]
\le
\frac{2\big(\mathbb{E}[f(\boldsymbol{\theta}^{0})] - f^*\big)}{\eta \mu_h T}
+ \frac{B_1}{\mu_h^2} + \frac{\eta L B_2}{\mu_h},
\end{equation}
}
where
\begin{align}
f^* &= \min_{\boldsymbol{\theta}} f(\boldsymbol{\theta}), \\
B_1 &= \frac{2\sigma_h^2(G^2 + \sigma_g^2)}{N} 
+ 2\mu_h^2(1-\gamma)(G^2 + \sigma_g^2), \\
B_2 &= (\mu_h^2 + \sigma_h^2)(G^2 + \sigma_g^2) + k\sigma_z^2.
\end{align}
\end{theorem}

\begin{IEEEproof}
See Appendix.
\end{IEEEproof}

\begin{remark}
Theorem~\ref{thm:convergence} indicates a convergence rate of $\mathcal{O}(1/T)$, which is limited by a steady-state error floor jointly governed by the compression ratio ($\gamma$), data heterogeneity ($\sigma_g^2$), and channel noise ($\sigma_h^2, \sigma_z^2$). 
Specifically, a lower compression ratio (smaller $\gamma$) increases the gradient bias via the $(1-\gamma)$ term in $B_1$. Higher data heterogeneity ($\sigma_g^2$) contributes to both $B_1$ and $B_2$, amplifying the steady-state error. 
Channel noise, particularly gain variance ($\sigma_h^2$) and noise power ($\sigma_z^2$), dominates $B_2$ and establishes the non-vanishing error floor. 
Consequently, achieving stable and efficient convergence requires reliable channels, moderate compression, and balanced data distribution in the OTA-FL system.
\end{remark}

\section{Experimental Results}

\subsection{Setup}
We evaluate AgeTop-$k$ on EMNIST \cite{cohen17emnist} and CIFAR-10 \cite{krizhevsky09learning} using a Convolutional Neural Network (CNN) \cite{lecun1998gradient} and ResNet-18 \cite{he16deepresidual}, respectively. 
Data are non-i.i.d. across $N=20$ clients following a symmetric Dirichlet distribution \cite{hsu19measuring} with $\alpha=0.3$. 
The wireless channel experiences Rayleigh fading with average gain $\mu_h = 1$. 
Learning rates are set to $\eta=10^{-4}$ (EMNIST) and $\eta=10^{-2}$ (CIFAR-10), with candidate ratio $\rho_r$ and transmission ratio $\rho_k$ ($r=\lfloor\rho_r d\rfloor$, $k=\lfloor\rho_k d\rfloor$). 
Experiments are performed using PyTorch on NVIDIA RTX 3090 GPU, with results averaged over 5 runs using different random seeds.

\subsection{System Performance}

\begin{figure}[t!]
    \centering
    \begin{subfigure}[t]{0.9\linewidth}
        \centering
        \includegraphics[width=\linewidth]{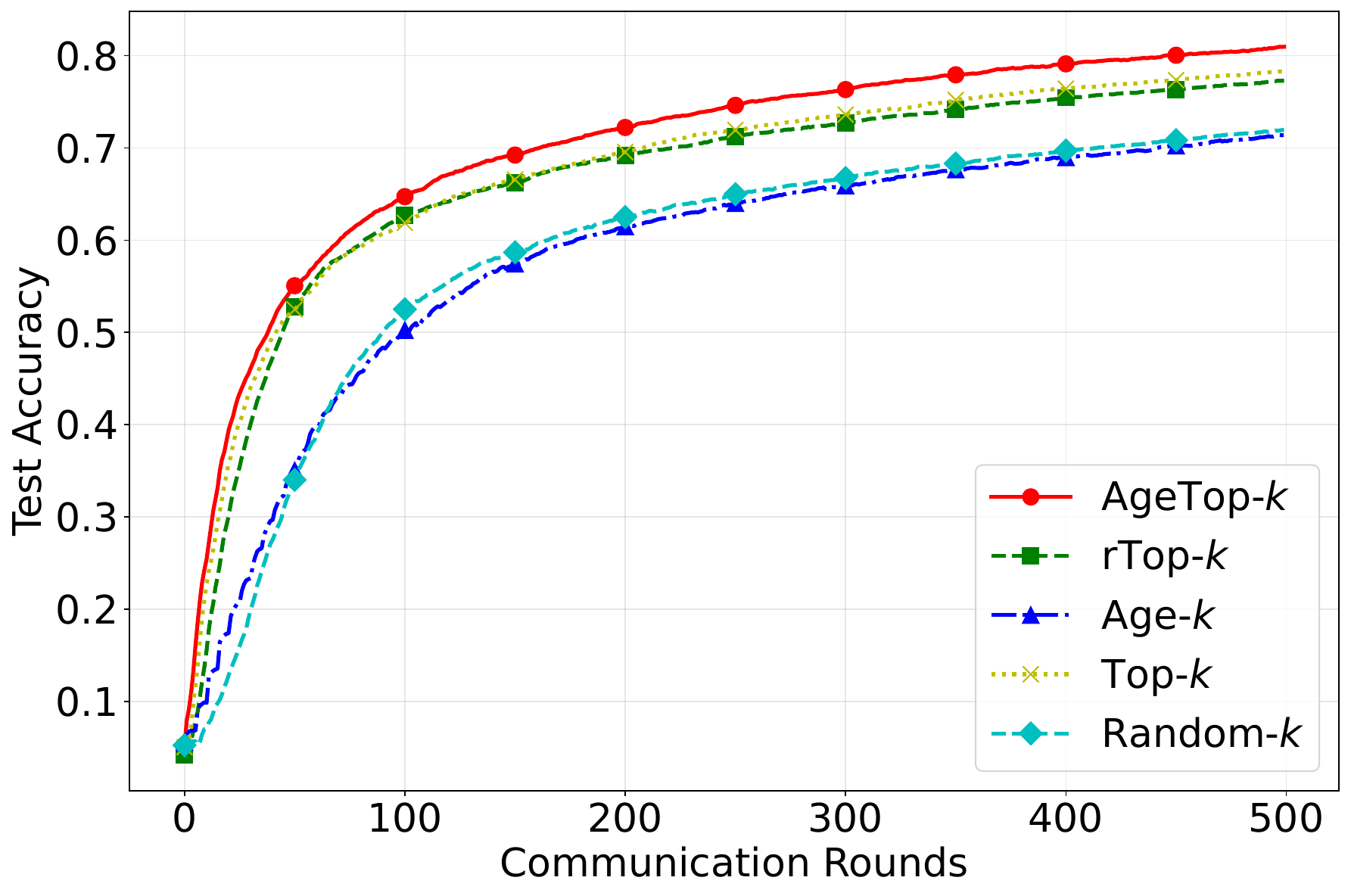}
        \caption{CNN on EMNIST ($\rho_r=0.3$, $\rho_k=0.2$)}
        \label{fig:mnist-main}
    \end{subfigure}
    
    \vspace{0.3cm}
    
    \begin{subfigure}[t]{0.9\linewidth}
        \centering
        \includegraphics[width=\linewidth]{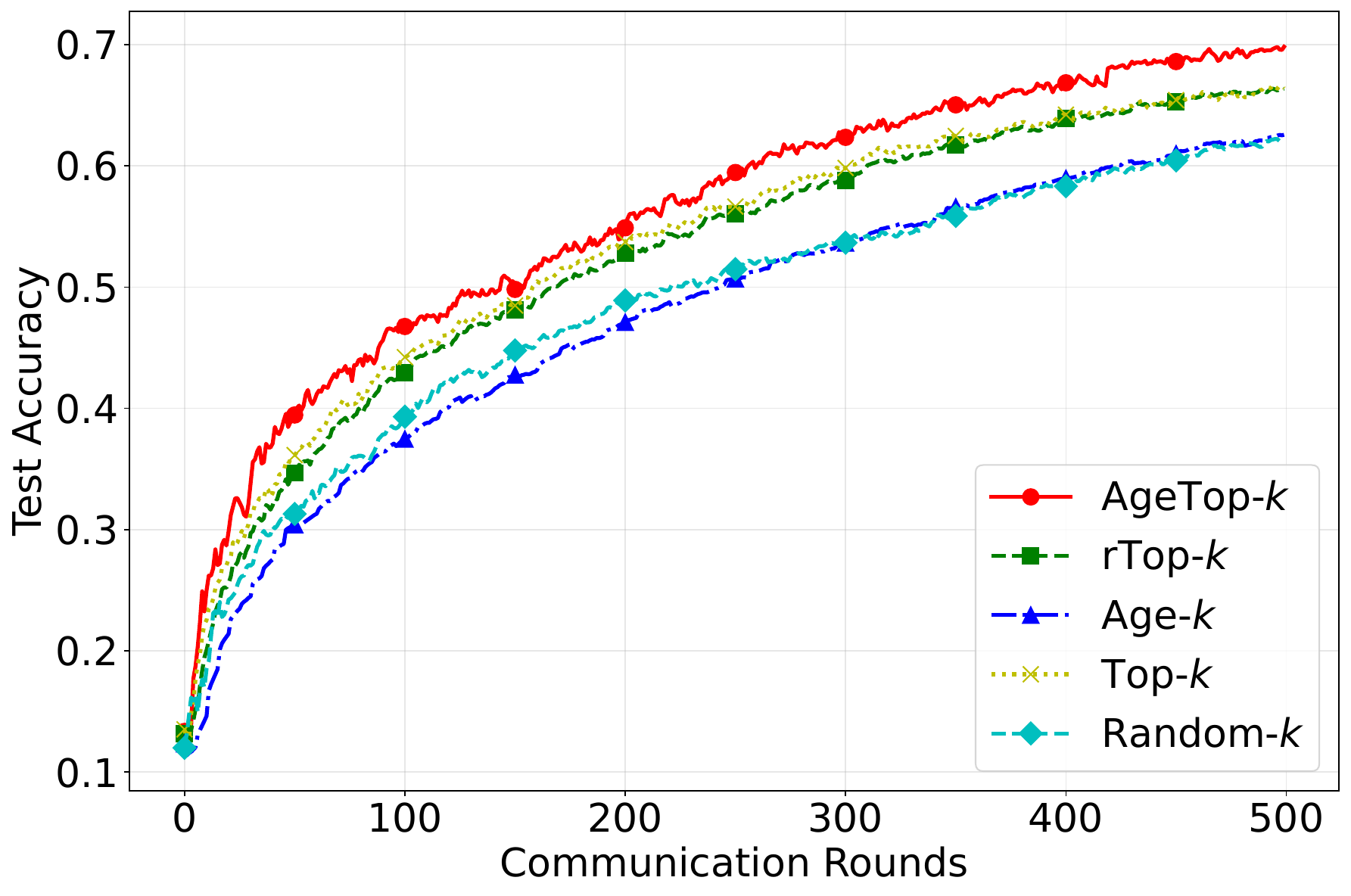}
        \caption{ResNet-18 on CIFAR-10 ($\rho_r=0.3$, $\rho_k=0.2$)}
        \label{fig:cifar-main}
    \end{subfigure}
    % \caption{Comparison of test accuracy between AgeTop-$k$ and baseline methods across different models and datasets.}
    \caption{Test accuracy comparison of AgeTop-$k$ and baseline methods across different models and datasets.}

    \label{fig:main-comparison}
\end{figure}

We compare AgeTop-$k$ with four baselines: Top-$k$, which chooses the $k$ largest-magnitude gradients; Random-$k$, which selects $k$ entries uniformly at random; Age-$k$, which selects the $k$ entries with the highest AoI ($\rho_r=1.0$); and rTop-$k$ \cite{barnes20rtop}, which randomly samples $k$ gradients from the top $r$ by magnitude.
As shown in Fig.~\ref{fig:main-comparison}, AgeTop-$k$ achieves the fastest convergence and highest test accuracy on both datasets.

Several insights can be drawn.
First, Age-$k$ performs comparably to Random-$k$, indicating that purely age-based selection is nearly as ineffective as random choice.
Second, Top-$k$ and rTop-$k$ show similar performance, implying that random sampling in rTop-$k$ offers little benefit over standard Top-$k$.
In contrast, AgeTop-$k$ outperforms all single-metric baselines, validating the advantage of the proposed two-stage selection strategy.
Moreover, its significant gain over rTop-$k$—which shares the same magnitude-based first stage—demonstrates that the AoI-guided second-stage selection is more effective and directional than random sampling.
By ensuring that neglected yet important gradients are timely updated, AgeTop-$k$ enhances update directionality and training efficiency, with negligible overhead, as the edge server only maintains a simple age vector.
Consistent gains across CNN and ResNet-18 further confirm the robustness of our method.

\begin{figure}[t!]
    \centering
    \includegraphics[width=0.9\linewidth]{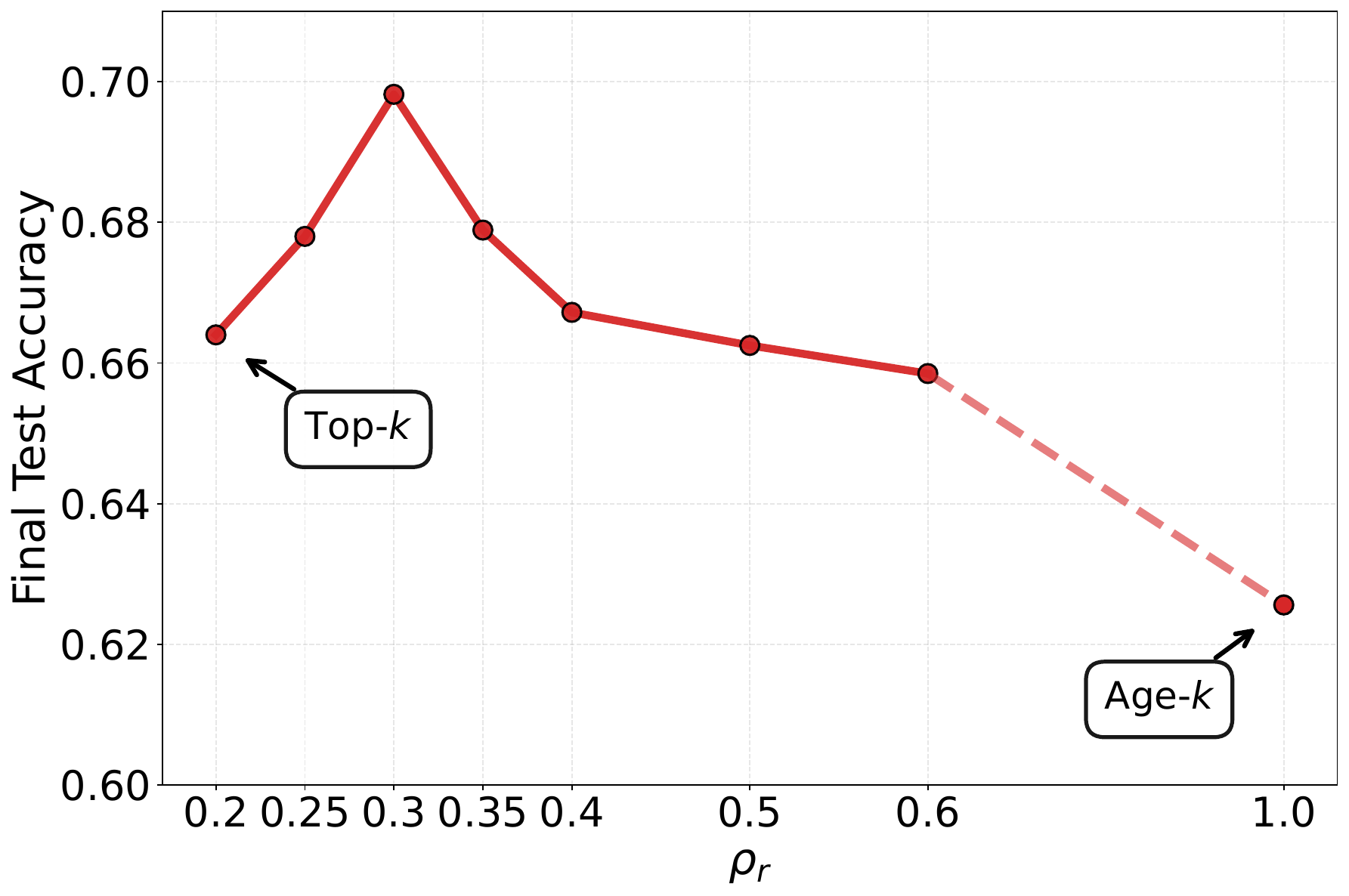} 
    % \caption{Impact of candidate ratio $\rho_r$ on the final test accuracy of AgeTop-$k$ with ResNet-18 on CIFAR-10 ($\rho_k=0.2$).}
    \caption{Effect of candidate ratio $\rho_r$ on the final test accuracy of AgeTop-$k$ with ResNet-18 on CIFAR-10 ($\rho_k=0.2$).}

    \label{fig:acc_r}
\end{figure}

As shown in Fig.~\ref{fig:acc_r}, we further analyze the impact of the candidate ratio $\rho_r$ under a fixed transmission ratio $\rho_k=0.2$.
The test accuracy first rises and then drops as $\rho_r$ increases, revealing a key trade-off.
Specifically, $\rho_r=0.2$ degenerates to standard Top-$k$, while $\rho_r=1.0$ corresponds to the purely age-based Age-$k$.
AgeTop-$k$ achieves the highest accuracy at $\rho_r=0.3$, demonstrating that combining gradient magnitude and AoI strikes an effective balance between exploiting recent updates and refreshing stale parameters.
The improvement from $\rho_r=0.2$ to $\rho_r=0.3$ underscores the benefit of incorporating AoI.
However, an excessively large $\rho_r$ introduces many low-magnitude gradients, reducing update efficiency.
These findings confirm that balancing gradient magnitude and AoI is essential for stable and efficient learning.

% \vspace{14mm}
\section{Conclusion}
In this work, we proposed AgeTop-$k$, an age-aware partial gradient update strategy for OTA-FL systems, aiming to alleviate the communication bottleneck imposed by the limited number of orthogonal waveforms. 
AgeTop-$k$ jointly considered the gradient magnitude and the timeliness of entry updates quantified by AoI to prioritize gradient sparsification, ensuring the timely transmission of stale yet significant gradient entries.
We derived the convergence rate of AgeTop-$k$ for non-convex smooth objective functions and validated its performance through extensive simulations.

% \vspace{-4mm}
\section{Appendix}
By the $L$-smoothness of $f(\boldsymbol{\theta}^{t})$, the expected loss at iteration $t+1$ can be bounded as follows:
{\small
\begin{align}
&\mathbb{E}\left[f\left(\boldsymbol{\theta}^{t+1}\right)\right] 
\leq\mathbb{E}\left[f\left(\boldsymbol{\theta}^{t}\right)\right]-\eta\mathbb{E}\left[\left\langle\nabla f\left(\boldsymbol{\theta}^{t}\right),\frac{1}{N} \sum_{n=1}^{N} h_n^{t}  \boldsymbol{S}^{t} \boldsymbol{g}_n^t\right\rangle\right] \nonumber\\
&\qquad \qquad \qquad \quad \!\!\!+\frac{\eta^{2}L}{2}\mathbb{E}\left[\left\|\frac{1}{N} \sum_{n=1}^{N} h_n^{t}  \boldsymbol{S}^{t} \boldsymbol{g}_n^t+(\hat{\boldsymbol{S}}^t)^\top \boldsymbol{\xi}^{t}\right\|_{2}^{2}\right] \nonumber\\
&=\mathbb{E}\left[f\left(\boldsymbol{\theta}^{t}\right)\right]\!
-\eta\mu_{h}\mathbb{E}\left[\left\|\nabla f\left(\boldsymbol{\theta}^{t}\right)\right\|_{2}^{2}\right] \nonumber\\
&\quad-\eta\mathbb{E}\left[\left\langle\nabla f\left(\boldsymbol{\theta}^{t}\right),\frac{1}{N} \sum_{n=1}^{N} h_n^{t}  \boldsymbol{S}^{t} \boldsymbol{g}_n^t 
-\mu_{h}\nabla f\left(\boldsymbol{\theta}^{t}\right)\right\rangle\right] \nonumber\\
&\quad+\!\frac{\eta^{2}L}{2}\mathbb{E}\left[\left\|\frac{1}{N} \sum_{n=1}^{N} h_n^{t}  \boldsymbol{S}^{t} \boldsymbol{g}_n^t+(\hat{\boldsymbol{S}}^t)^\top \boldsymbol{\xi}^{t} \right\|_{2}^{2}\right]\nonumber\\
&\leq\mathbb{E}\left[f\left(\boldsymbol{\theta}^{t}\right)\right]
-\frac{\eta\mu_{h}}{2}\mathbb{E}\left[\left\|\nabla f\left(\boldsymbol{\theta}^{t}\right)\right\|_{2}^{2}\right] \nonumber\\
&\quad+\frac{\eta}{2\mu_{h}}\underbrace {\mathbb{E}\left[\left\|
\frac{1}{N} \sum_{n=1}^{N}h_n^{t}  \boldsymbol{S}^{t} \boldsymbol{g}_n^t 
-\mu_{h}\nabla f\left(\boldsymbol{\theta}^{t}\right)\right\|_{2}^{2}\right]}_{Q1}\nonumber\\
&\quad+\frac{\eta^{2}L}{2}\underbrace{\mathbb{E}\left[\left\|
\frac{1}{N}\sum_{n=1}^{N}h_n^{t} \boldsymbol{S}^{t} \boldsymbol{g}_n^t
+(\hat{\boldsymbol{S}}^t)^\top \boldsymbol{\xi}^{t} \right\|_{2}^{2}\right]}_{Q2}.
\end{align}
}
According to Assumption 1, $Q_{1}$ can be bounded as
{\small
\begin{align}
&Q_{1}
= \mathbb{E}\Bigg[\Bigg\|
\frac{1}{N}\sum_{n=1}^{N} h_n^{t} \boldsymbol{S}^{t}\boldsymbol{g}_n^t
- \frac{1}{N}\sum_{n=1}^{N} \mu_{h}\boldsymbol{S}^{t}\boldsymbol{g}_n^t \nonumber \\
& \quad\quad\quad+ \frac{1}{N}\sum_{n=1}^{N} \mu_{h}\boldsymbol{S}^{t}\boldsymbol{g}_n^t
- \frac{1}{N}\sum_{n=1}^{N} \mu_{h}\boldsymbol{g}_n^t
\Bigg\|_{2}^{2}\Bigg]\nonumber \\
&\leq \!2\mathbb{E}\!\!\left[\!\left\|\frac{1}{N}\!\!\sum_{n=1}^{N}\!\!\left(h_n^t\!-\!\mu_h\right)\!\boldsymbol{S}^{t}\!\boldsymbol{g}_n^t\right\|_2^2\right]\!\!
\!+ \!2\mathbb{E}\!\!\left[\left\|\!\frac{1}{N}\!\!\sum_{n=1}^{N}\!\mu_h\!\!\left(\boldsymbol{S}^{t}\!\boldsymbol{g}_n^t\!\!\!-\!\boldsymbol{g}_n^t\right)\right\|_2^2\right]\nonumber \\
&\leq \frac{2\sigma_h^2}{N^2}\sum_{n=1}^{N}\mathbb{E}\left[\left\|\boldsymbol{S}^{t}\boldsymbol{g}_n^t\right\|_2^2\right]
+ \frac{2\mu_h^2}{N}\sum_{n=1}^{N}\mathbb{E}\left[\left\|\boldsymbol{S}^{t}\boldsymbol{g}_n^t-\boldsymbol{g}_n^t\right\|_2^2\right]\nonumber \\
&\leq \frac{2\sigma_h^2}{N^2}\sum_{n=1}^{N}\mathbb{E}\left[\left\|\boldsymbol{g}_n^t\right\|_2^2\right]
+ \frac{2\mu_h^2(1-\gamma)}{N}\sum_{n=1}^{N}\mathbb{E}\left[\left\|\boldsymbol{g}_n^t\right\|_2^2\right].
\end{align}}Subsequently, we apply Assumption 2 and Assumption 3 to obtain
{\small
\begin{align}
\frac{1}{N}\sum_{n=1}^{N}\mathbb{E}
\left[\left\|\boldsymbol{g}_n^t\right\|_{2}^{2}\right]
&=\frac{1}{N}\sum_{n=1}^{N}\mathbb{E}
[\left\|\boldsymbol{g}_n^t-\nabla f(\boldsymbol{\theta}^{t})
\right\|_{2}^{2}+\left\|\nabla f(\boldsymbol{\theta}^{t})\right\|_{2}^{2}]\nonumber\\
&\leq \sigma_{g}^{2} + G^{2}.
\end{align}}Substituting (24) into $Q_{1}$, $Q_{1}$ can be bounded as
{\small
\begin{align}
Q_{1}
& \leq \frac{2\sigma_h^2(G^2+\sigma_g^2)}{N} + 2\mu_h^2(1-\gamma)(G^2+\sigma_g^2).
\end{align}
}For convenience, we define the above expression as $B_{1}$. 
Noting that $\mathbb{E}[\boldsymbol{\xi}^{t}]=0$, we can expand $Q_{2}$ as follows:
{\small
\begin{align}
Q_{2}
&= \mathbb{E}\left[\left\|
\frac{1}{N}\sum_{n=1}^{N}h_n^{t} \boldsymbol{S}^{t} \boldsymbol{g}_n^t
\right\|_{2}^{2}\right]
+\mathbb{E}\left[\left\|(\hat{\boldsymbol{S}}^t)^\top \boldsymbol{\xi}^{t}\right\|_{2}^{2}\right]\nonumber\\
&\leq \frac{\mu_h^2 + \sigma_h^2}{N} \sum_{n=1}^{N}\mathbb{E} 
\left[\left\|
 \boldsymbol{S}^{t} \boldsymbol{g}_n^t
\right\|_{2}^{2}\right]
+k\sigma_{z}^{2}\nonumber\\
&\leq \frac{\mu_h^2 + \sigma_h^2}{N} 
\sum_{n=1}^{N}\mathbb{E} 
\left[\left\|
\boldsymbol{g}_n^t
\right\|_{2}^{2}\right]
+k\sigma_{z}^{2}.
\end{align}}Substituting (24) into $Q_{2}$, $Q_{2}$ can be bounded as
{\small
\begin{align}
Q_{2}
&\leq (\mu_h^2+\sigma_h^2)(\sigma_{g}^2 + G^2) + k\sigma_z^2.
\end{align}}For convenience, we define the above expression as $B_{2}$. 
Substituting $B_{1}$ and $B_{2}$ into (22), we can obtain
{\small
\begin{align}
\mathbb{E}\left[f\left(\boldsymbol{\theta}^{t+1}\right)\right]
&\leq\mathbb{E}\left[f\left(\boldsymbol{\theta}^{t}\right)\right]-\frac{\eta\mu_h}{2}\mathbb{E}\left[\left\|\nabla f\left(\boldsymbol{\theta}^{t}\right)\right\|_{2}^{2}\right] \nonumber\\
&\quad+\frac{\eta}{2\mu_h}B_1 +\frac{\eta^2L}{2}B_2. 
\end{align}}Finally, by induction we reach:
{\small
\begin{align}
% &\frac{1}{T}\sum_{t=0}^{T-1}\mathbb{E}\left[\left\|\nabla f\left(\boldsymbol{\theta}^{t}\right)\right\|_{2}^{2}\right] 
% \leq \frac{2}{\eta T}\left(\mathbb{E}\left[f(\boldsymbol{\theta}^{0})\right]-f^*\right)+\eta LB_{1}+B_{2}.
\frac{1}{T}\sum_{t=0}^{T-1}\mathbb{E}\!\left[\left\|\nabla f\!\left(\boldsymbol{\theta}^{t}\right)\right\|_{2}^{2}\right]
&\leq
\frac{2\!\left(\mathbb{E}\!\left[f\!\left(\boldsymbol{\theta}^{0}\right)\right]-f^*\right)}{\eta\mu_h T}
+\frac{B_1}{\mu_h^{2}}+\frac{\eta L B_2}{\mu_h}.
\end{align}}The proof is completed.

\bibliographystyle{IEEEtran}
\bibliography{IEEEabrv,references}
\end{document}